\pdfoutput=1
\documentclass[twocolumn,twocolappendix]{aastex631}

\usepackage{amsmath}

\shorttitle{Observation bias in GW detections}
\shortauthors{Veske et al.}

\begin{document}

\title{Characterizing the observation bias in gravitational-wave detections and finding structured population properties}

\author[0000-0003-4225-0895]{Do\u{g}a Veske}
\correspondingauthor{Do\u{g}a Veske}
\email{dv2397@columbia.edu}
\affiliation{Department of Physics, Columbia University in the City of New York, New York, NY 10027, USA}
\author[0000-0001-5607-3637]{Imre Bartos}
\affiliation{Department of Physics, University of Florida, PO Box 118440, Gainesville, FL 32611-8440, USA}
\author[0000-0003-1306-5260]{Zsuzsa M\'arka}
\affiliation{Columbia Astrophysics Laboratory, Columbia University in the City of New York, New York, NY 10027, USA}
\author[0000-0002-3957-1324]{Szabolcs M\'arka}
\affiliation{Department of Physics, Columbia University in the City of New York, New York, NY 10027, USA}

\begin{abstract}

The observed distributions of the source properties from gravitational-wave detections are biased due to the selection effects and detection criteria in the detections, analogous to the Malmquist bias. In this work, this observation bias is investigated through its fundamental statistical and physical origins. An efficient semi-analytical formulation for its estimation is derived which is as accurate as the standard method of numerical simulations, with only a millionth of the computational cost. Then, the estimated bias is used for unmodelled inferences on the binary black hole population. These inferences show additional structures, specifically two peaks in the joint mass distribution around binary masses $\sim10$ M$_\odot$ and $\sim30$ M$_\odot$. Example ready-to-use scripts and some produced datasets for this method are shared in an online repository.
\end{abstract}
\keywords{Gravitational wave sources (677); Astrophysical black holes (98)}

\section{Introduction}

The vast majority of gravitational-wave detections, including the first detection \citep{PhysRevLett.116.061102}, have been from binary black hole (BBH) mergers \citep{Abbott_2019_1,abbott2021gwtc2}. With the increasing number of detected BBH pairs from mergers, inferences on the population of BBHs and their formation channels have been made \citep{Abbott_2019,theligoscientificcollaboration2021population}. Such analyses shed light on the origins of binary black holes \citep{Zevin_2021}, providing hints on stellar evolution \citep{PhysRevD.93.084029}, pair-instability mass gap \citep{Woosley_2017,woosley2021pairinstability}, hierarchical mergers \citep{PhysRevLett.123.181101,Gayathri_2020,10.1093/mnrasl/slaa123,Veske_2021,kimball2020evidence}, primordial black holes \citep{PhysRevLett.116.201301,CLESSE2017142} and other exotic objects \citep{PhysRevLett.126.081101}. For the accurate estimates of the actual populations and also for the other interpretations based on observations, the effect of population parameters on the observation should be correctly understood as the observed population parameters can have an observation bias; i.e. the observed relative fraction of the events which are detected easier will be higher than their actual astrophysical relative fraction, analogously to the Malmquist bias \citep{1922MeLuF.100....1M}.
Usually, such biasing effects on the observations are accounted with numerical simulation campaigns where a large set of BBH mergers' waveforms according to a population model are numerically simulated. Then the simulated mergers are tried to be detected similarly to the real detection pipelines either in the presence of a noise similar to the actual detectors’ noise or, for more accurate estimations, they are 'injected' into real data segments, which do not contain confirmed detections, and are tried to be detected in that setting \citep{abbott2021gwtc2}. The difference between the simulated and detected populations characterizes the observation bias \citep{Mandel_2019,theligoscientificcollaboration2021population}. Early investigation of this problem under certain simplifying conditions was made in \cite{PhysRevD.47.2198}. Recently, this problem is being attacked with new data processing techniques such as neural networks and machine learning as well \citep{2020arXiv201201317T,Gerosa:2020pgy,gerardi2021unbiased}.

In this letter, that observation bias is analyzed semi-analytically with the aim of devising a computationally much less expensive way for finding the bias while also providing physical intuitions on it. The accuracy of the method is verified by the traditional ways of computing it via simulations. Of course the estimated result cannot be expected to be as accurate as the result of the real data injections; since the rationale of the real data injections is the fact that the behavior of the detectors' noise is not understood fully and hence could not have been modelled very accurately. However, for studies not requiring extreme accuracy or for which such an accuracy is not feasible due to a large parameter space, such as when estimating the expected observed distribution from a certain formation channel with an uncertain astrophysical distribution \citep{Veske_2021}, accounting the observation bias in cosmological estimations from GWs \citep{PhysRevD.100.103523} or studying a future detector \citep{10.1093/mnras/sty3321}; having a sufficiently accurate, easy and computationally cheap method for accounting the observation bias may be very useful for the whole scientific community, from theorists to detector designers. Finally, the method is applied on the BBH mergers in the gravitational wave transient catalogs GWTC-1 \citep{Abbott_2019_1} and GWTC-2 \citep{abbott2021gwtc2} to find unmodelled inferences for the representative black hole population. The current aim of these unmodelled estimates is mainly to see the possible underlying structures in the population properties which are not parametrized by the current models. As more BBH mergers are observed, models with simple parametrizations fail to explain the observed population and hence more complex models with many parameters have started to be used \citep{theligoscientificcollaboration2021population}. Identification of these structures can guide the development of new parametrizations instead of blindly guessing the distributions with commonly used mathematical functions. Moreover, in the future as there are more detections, the need for modelled inferences may disappear and they may be replaced by unmodelled inferences since modelled inferences are essentially used due to their robustness against statistical fluctuations. This initial study is limited to the mergers of non-spinning quasicircular BBHs observed by interferometric gravitational-wave detectors via conventional matched filtering \citep{leon} in the presence of an additive Gaussian noise. First, fundamentals of the bias are explained in Sec. \ref{sec:understand} referring to the statistics and physics behind it. In Sec. \ref{sec:finding}, the effects of the bias on the observed mass distributions are calculated semi-analytically with a list of numerically generated signal-to-noise ratios in the detectors for different masses. In Sec. \ref{sec:cosmo} the bias is used to infer structures of the astrophysical distributions from detections. We summarize and conclude in Sec. \ref{sec:conc}.

\section{Understanding the bias}
\label{sec:understand}
Interferometric gravitational-wave detectors are designed to measure the variations in the lengths of the arms of them. They are very sensitive position detectors and the signal power measured by them via matched filtering in the presence of a white noise is proportional to the square of the distance difference between the ends of their arms. This methodology intrinsically differs from most of other astronomical detections where a fraction of the radiated energy in an event is directly detected, generally via excitation of electrons in a semiconductor device or a crystal through the absorption of the received energy. Whereas, as one would expect from non-relativistic classical physics, the physical power deposited to an interferometric gravitational-wave detector at rest is proportional to the square of the induced oscillation speed to the free ends of the arms. This non-proportionality between the signal energy and the absorbed physical energy in gravitational-wave detection demonstrates a non-trivial observation bias where not necessarily events with high emitted energy are favored in the detection.
\subsection{Origin of the bias}

The observational bias essentially depends on the signal power generated by a physical configuration and the noise power present in the detector.
The configurations which generate a higher signal to noise power ratio (SNR)\footnote{The power SNR ($\rho^2$) is the square of the amplitude SNR, which is defined in \cite{Allen_2012}, and is the additive quantity for a network of detectors.} are easier to be observed and consequently the relative fraction of observed sources become biased in favor of those which generate a higher SNR. In this letter the physical configuration for a BBH of interest includes the source frame masses of the heavy and light black holes ($m_1$, $m_2$ respectively), the luminosity distance between the BBH and the detector ($r$), the corresponding cosmological redshift at that distance ($z(r)$), angular location of the BBH on the sky ($\mathbf{\Omega}$), the inclination angle of the binary's orbital angular momentum to the line of sight ($\iota$) and the polarization angle ($\psi$) which is the angle between the x-y coordinates of the detector frame and radiation frame which varies with the orientation of the orbital angular momentum around the direction of the line of sight. The black holes are considered to be non-spinning and consequently the binary systems are not precessing. For simplicity the variations in the signal power induced by the initial orbital phase of the binary, which has the only effect of shifting the oscillatory waveform in the envelope of the waveform for non-precessing systems, are neglected. Such an effect on the signal power is expected to be on the order of few percent maximum. If one desires to be more accurate, the signal power for each configuration can be averaged uniformly over the initial orbital phase of the binary as well although this approximation was found to be a subdominant source of error in the described method in this letter.
The main properties of the BBHs which are affected by an observation bias investigated in this letter are the masses of the involved black holes and the distance of BBHs; such as the joint or marginalized distributions of the masses $P(m_1,m_2)$ or $P(m_1)$, or the evolution of the BBH merger rate which is related to the distance distribution $P(r)$.

Since the power generated in the gravitational-wave detectors are dependent on the masses, observed mass distributions are different than the actual distributions; i.e. $P(m_1|D)\neq P(m_1)$ where $D$ is used to denote the events being detected. Below, $m_1$ is used to demonstrate the relations between the observed and actual distributions. Similar relations can be written for any property along these lines. The relation between the actual and observed distributions can be written by using the Bayes' rule as
\begin{equation}
    P(m_1|D)=\frac{P(D|m_1)P(m_1)}{P(D)}
\end{equation}
$P(D|m_1)$ can be further expanded as 
\begin{multline}
    P(D|m_1)=\int P(D|m_1, m_2, r, \mathbf{\Omega}, \iota) P(m_2|m_1) \\ \times P(\mathbf{\Omega}) P(r) P(\iota)P(\psi) dm_2dr d\mathbf{\Omega}d\iota d\psi
    \label{eq:bayes}
\end{multline}
where the extrinsic properties are considered to be independent of all the other properties and $m_2$ is considered to be dependent on $m_1$ since there is at least one dependency of $m_2\leq m_1$ by their definition.
Denoting the average power SNR generated in the detector from certain intrinsic and extrinsic properties with $E$, and neglecting the difference between the expected and observed SNR due to noise fluctuations (considering the average SNR as the deterministic SNR value for fixed intrinsic and extrinsic properties), the detection likelihood can be written as
\begin{multline}
    P(D|m_1, m_2, r, \mathbf{\Omega}, \iota,\psi)\\=P(E(m_1, m_2, r, \mathbf{\Omega}, \iota,\psi)>\rho^2_{th}) \\ =\Theta(E(m_1, m_2, r, \mathbf{\Omega}, \iota,\psi)-\rho^2_{th})
    \label{eq:integral}
\end{multline}
where $\Theta$ is the Heaviside step function. If the difference between the expected and observed SNR were not neglected, a smoothly increasing function around $\rho^2$ from 0 to 1 (similar to the error function) would be used instead of the step function \citep{thrane_talbot_2020}. 

Dependency of $E$ to extrinsic properties can be calculated analytically whereas the dependency on mass cannot be found exactly due to complete gravitational-waveforms being non-analytical (see Appendix \ref{sec:gw}). Only the inspiral and ringdown phases of the waveforms have analytical forms without an analytical solution for the merger phase. In order to understand the full dependency of SNR on the mass, in the next section, numerical computations were performed which takes in account the detectors' different sensitivities at different frequencies.

\subsection{Mass dependency of SNR}
\label{sec:mass}

As mentioned in the previous section, the exact mass dependency of the SNR cannot be found analytically. Even the contribution from the merger phases which has an analytical solution cannot be determined as the SNR is proportional to the integral of the amplitude square of the wave (see Appendix \ref{sec:mf}). Although the integrand is analytical, integration limits are not since the next merger phase is non-analytical. Due to the overall non-analytic behavior of the SNR with masses, the dependency is investigated empirically by computing the generated SNR over a range of mass combinations with fixed extrinsic properties. The waveforms are generated by using the NRHybSur3dq8 surrogate waveform model \citep{PhysRevD.99.064045} via the \textsc{gwsurrogate} package \citep{PhysRevX.4.031006}. In order to make comparisons with the results from traditional simulation studies, the detectors' noise the power spectral densities (PSD) are chosen as aLIGOMidLowSensitivityP1200087 which is the PSD used in the simulation studies of gravitational-waves (i.e. via \textsc{ligo.skymap} package\footnote{\url{https://lscsoft.docs.ligo.org/ligo.skymap/}}) used for representing a pessimistic sensitivity estimate for the LIGO detectors \citep{2015} during their third observing run O3. The power SNR was calculated via matched filtering (via Eq. \eqref{eq:mf1}) for every pair of integer valued black hole masses in [10,100]M$_\odot\times$[10,100]M$_\odot$. The templates used in real searches performed by LIGO Scientific and Virgo Collaborations include only the dominant wave mode (2,2) and have a low frequency cut at 15 Hz \citep{canton2017designing,PhysRevD.95.044028,PhysRevD.95.104045,PhysRevD.99.024048,abbott2021gwtc2}. In order to be as realistic as much as possible, here the waveform templates were also assumed to be alike while the astrophysical signals were considered to have all the available wave modes\footnote{Considering the available modes in the NRHybSur3dq waveform which are (2,2), (2,1), (2,0), (3,3), (3,2), (3,1), (3,0), (4,4), (4,3), (4,2) and (5,5)} above 15 Hz. Due to the mismatches between the templates and astrophysical signals for same parameters, the highest SNR generating template may not have the same parameters with the astrophysical waveform \citep{PhysRevD.93.084019}. In order to have the (5,5) mode, and other modes, completely above 15 Hz; waveforms were generated from where (2,2) mode reaches 6 Hz.

\begin{figure}
\gridline{
    \fig{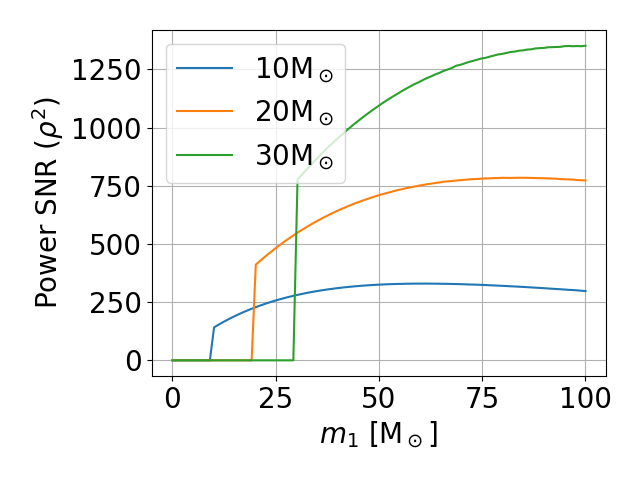}{0.24\textwidth}{(a)}
    \fig{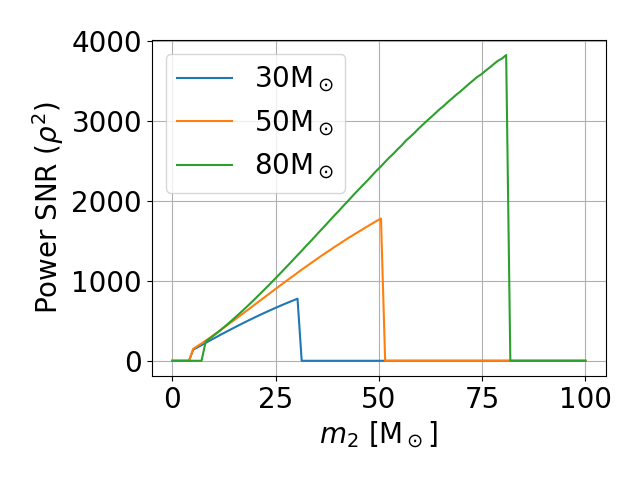}{0.24\textwidth}{(b)}}
    \gridline{
    \fig{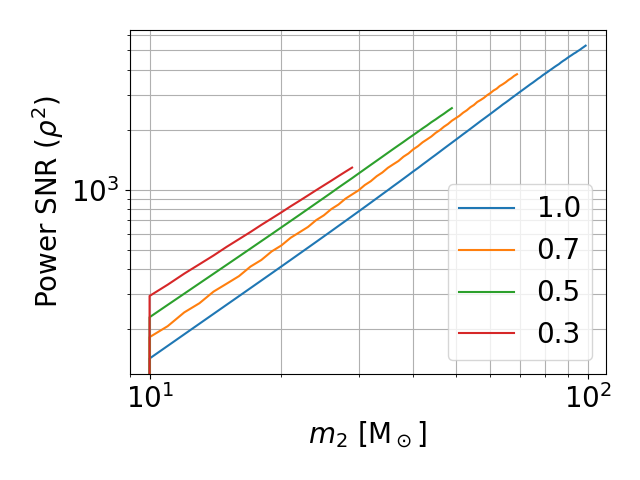}{0.24\textwidth}{(c)}
    \fig{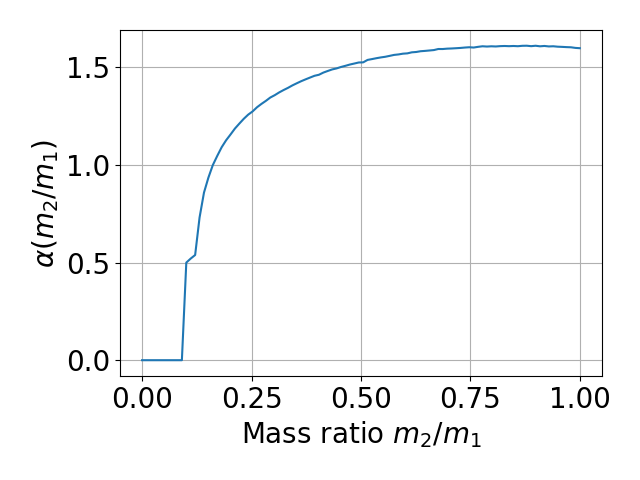}{0.24\textwidth}{(d)}
}

\caption{Dependencies of power SNR for different mass configurations: (a) $m_1$ dependency of SNR for constant $m_2$; (b) $m_2$ dependency of SNR for constant $m_1$; (c) the power law dependency of SNR on the masses for select mass ratios, as lines in log-log scale; (d) exponent of the total mass dependency of SNR for fixed mass ratios. The SNRs were calculated for a face-on BBH mergers at 1 Gpc in the absence of cosmological redshift considering one of the polarizations with a unit antenna factor. The jitter in SNR graphs at 1\% level was found to be due to the discretized non-trivial noise spectrum.}
\label{fig:SNR}
\end{figure}

It was observed that the SNR increases approximately linearly with the mass of the smaller black hole for a constant heavier mass. 
A similar dependency is present in the emitted gravitational-wave energy as well\footnote{The emitted energy is in $m_2$[9.5\%,12\%] for mass ratio $m_1/m_2$ in [1,9] \citep{Barausse_2012}.} although there needs not be a direct correspondence as mentioned before. On the other hand, SNR varies non-trivially with the heavier mass $m_1$ for a constant small mass $m_2$. 
The variation with the heavier mass has a sublinear increase at the start which eventually becomes a stall and then a decrease at extreme mass ratios. 
The variations of the SNR with smaller and heavier masses when the other one is constant are given in Fig. \ref{fig:SNR} for select masses. The final analyzed dependency is on the total mass. For a constant mass ratio, SNR may be expected to increase with $(m_1+m_2)^{5/2}$ (see Eq. \eqref{eq:h2}). Although there are several non-analytical complications, the SNR is nevertheless found to be fit very well by a power of the total mass for a constant mass ratio $\rho^2\propto(m_1+m_2)^{\alpha(m_2/m_1)}$, where the exponent $\alpha(m_2/m_1)$ is a function of the mass ratio. The power law dependency of the SNR for constant mass ratio and empirically found $\alpha(m_2/m_1)$ are shown in Fig. \ref{fig:SNR}. This relation is needed when determining the bias in the presence of a cosmological redshift where the observed mass ratio remains unchanged but the observed total mass is amplified. The practical applicability of this empirical relation will further be verified in the next section by a comparison against simulations.

\section{Finding the bias and estimating observed distributions}
\label{sec:finding}
In the previous section the basics for estimating the bias on the properties of the BBH mergers were laid down. In this section, the effect of the bias is computed and observed distributions are estimated. First a homogeneous universe without the cosmological redshift is considered and then redshift and changing source density will also be included. Accuracy of the obtained distributions are verified via simulated injections. Such simulations are currently the used method for accounting the observation bias.

\subsection{Static and homogeneous universe}
When there is no expansion hence cosmological redshift, the power SNR generated in the network of $N_d$ detectors can be written by decoupling dependency of several properties
\begin{equation}
    E(m_1,m_2,r,\mathbf{\Omega},\iota)=\frac{\sum_{i=1}^{N_d}E_{0,i}(m_1,m_2)f_i(\mathbf{\Omega},\iota,\psi)}{r^2/r_0^2}
\end{equation}
where, for the dominant (2,2) wave mode, $f$ is defined as $f(\mathbf{\Omega},\iota,\psi)=(F_+^2(\mathbf{\Omega},\psi)(\frac{1+cos^2\iota}{2})^2+F_\times^2(\mathbf{\Omega},\psi)cos^2\iota)$, $F_+$ and $F_\times$ are the antenna patterns of the detectors for two tensor polarizations, and $E_0(m_1,m_2)$ is the power SNR generated by masses $m_1$ and $m_2$ when $f=1$ at a distance $r_0$ in the absence of cosmological redshift, and sum over $i$ represents different detectors. Although the $\iota$ dependency of each wave mode of order $|m|$ is different; in this letter the effect of the inclination angle $\iota$ is carried as if there are only the $|m|=2$ modes which is the case for the search templates but not the astrophysical signals. This approximation assumes the relative mismatch between the templates and real signals to be independent of $\iota$. Therefore the SNR estimates may have up to $\sim$10\% error, especially for extreme mass ratio and high-inclination binaries. When there is no redshift in an homogeneous universe, the distribution of $r$ is $P(r)=3r^2/r_{max}^3$ for $r<r_{max}$ where $r_{max}$ is well beyond the observation horizon of the detector network and the detection likelihood of $m_1$ can be written as

\begin{multline}
    P(D|m_1)=\int \Theta(\frac{\sum_{i=1}^{N_d}E_{0,i}(m_1,m_2)f_i(\mathbf{\Omega},\iota,\psi)}{r^2/r_0^2}-\rho^2_{th})\\ \times P(m_2|m_1)
    \frac{3r^2}{r_{max}^3}P(\mathbf{\Omega})P(\iota)P(\psi)dm_2d\mathbf{\Omega}d\iota d\psi dr\\
    = \int_{\sqrt{\sum_{i}r_0^2E_{0,i}(m_1,m_2)f_i(\mathbf{\Omega},\iota,\psi)\rho^{-2}_{th}}>r} P(m_2|m_1)\frac{3r^2}{r_{max}^3}\\ \times P(\mathbf{\Omega}) P(\iota)P(\psi)dm_2d\mathbf{\Omega}d\iota d\psi dr \\
    = \int (\sum_ir_0^2E_{0,i}(m_1,m_2)f_i(\mathbf{\Omega},\iota,\psi)\rho^{-2}_{th}r_{max}^{-2})^{3/2}\\ \times P(m_2|m_1) P(\mathbf{\Omega})P(\iota)P(\psi)dm_2d\mathbf{\Omega}d\iota d\psi
    \\=\frac{r_0^3}{r_{max}^3\rho^3_{th}}\int (\sum_iE_{0,i}(m_1,m_2)f_i(\mathbf{\Omega},\iota,\psi))^{3/2} P(m_2|m_1)\\ \times P(\mathbf{\Omega})P(\iota)P(\psi)dm_2d\mathbf{\Omega}d\iota d\psi
\end{multline}

Interestingly, if the frequency sensitivities of the detectors are proportional to each other, i.e. $E_{0,1}=c_1E_{0,2}$ and if the merger rate is constant, then neither the distributions of $\mathbf{\Omega}$ and $\iota$ nor the $f$ function do not change the $m_1$ dependency of the result of Eq. \eqref{eq:integral}. They only bring an overall factor which is eventually cancelled with the normalization constant in Eq. \eqref{eq:bayes}. In this case, $E_0$ can be factored out from the sum and the detection likelihood becomes
\begin{multline}
    P(D|m_1)= \int E_0(m_1,m_2)^{3/2}P(m_2|m_1)dm_2\\ \times \frac{r_0^3}{r_{max}^3\rho^3_{th}}\int (\sum_ic_if_i(\mathbf{\Omega},\iota,\psi))^{3/2}P(\mathbf{\Omega})P(\iota)P(\psi)d\mathbf{\Omega}d\iota d\psi
    \label{eq:6}
\end{multline}
The factor in the second line of Eq. \eqref{eq:6} is cancelled with the normalization in the denominator in Eq. \eqref{eq:bayes} since it does not depend on  $m_1$. The observational bias on $m_1$ becomes proportional to $\int E_0(m_1,m_2)^{3/2}P(m_2|m_1)dm_2$ and consequently the observed distribution can be written as 
\begin{equation}
    P(m_1|D)=\frac{P(m_1)\int E_0(m_1,m_2)^{3/2}P(m_2|m_1)dm_2}{\int P(m_1) E_0(m_1,m_2)^{3/2}P(m_2|m_1)dm_2dm_1}
    \label{eq:biasm1}
\end{equation}
When the frequency sensitivities of the detectors are proportional to each other, neither the antenna factors, distribution of the sources in the sky, the distribution of the inclination of the orbits of BBHs nor the detection threshold on SNR affect the observed distribution of $m_1$ when there is no cosmological redshift. In reality with cosmological redshift, this simple calculation is appropriate for use with high detection thresholds or with weak detectors, where the horizon of the search is at low redshifts. Likewise, it can be used for searches of less powerful sources such as binary neutron stars. With current detectors; LIGO \citep{2015} Hanford is $\sim 1.5$ and Virgo \citep{Acernese_2014} is $\sim 6$ times less sensitive than LIGO Livingston (from their noise power spectral densities around 100Hz\footnote{From \url{https://www.gw-openscience.org/detector_status/}}) with a similar frequency dependency. Therefore this approximation can be used for them for more crude estimations.

In order to demonstrate the accuracy of this estimation, a simulation study using the \textsc{LALApps} and \textsc{ligo.skymap} packages was done, injecting a population of BBH mergers in the absence of a cosmological redshift with uniformly distributed masses in the $(m_1,m_2)$=[10,100]M$_\odot\times$[10,100]M$_\odot$ space. The orbital orientation of the BBHs and their position in volume were uniformly randomized with a maximum luminosity distance of 10 Gpc which is beyond the maximum detection distance for the chosen detector configuration for a 100M$_\odot$+100M$_\odot$ binary. The local rate density of the mergers was assumed to be constant. The mergers were detected with two LIGO detectors with the same PSD used to compute the SNRs via surrogate waveforms (aLIGOMidLowSensitivityP1200087) with a detection threshold of $\rho^2_{th}=144$ on the network power SNR. IMRPhenomPv2 waveforms were used. Furthermore, an estimation by assuming only the dominant (2,2) mode in the astrophysical signals is done which actually is more meaningful for a comparison with the simulations since the astrophysical waveforms used in the simulations include only the dominant mode. The histogram of $m_1$ values for the injected and detected BBHs overlaid  with the estimation using Eq. \eqref{eq:biasm1} and the SNR distribution used in Sec. \ref{sec:mass} can be seen in Fig \ref{fig:hist2}. It is seen that the observed distribution is accurately estimated. The presence of higher order modes doesn't show a meaningful effect on the estimation.

\subsection{Expanding universe}
\label{sec:exp}
The cosmological redshift $1+z(r)$ modifies the received gravitational-waveform as if the masses are multiplied by $1+z(r)$. Therefore the SNR generated as a function of BBH properties can be written as

\begin{multline}
    E(m_1,m_2,r,\mathbf{\Omega},\iota)\\=\frac{\sum_iE_{0,i}(m_1(1+z(r)),m_2(1+z(r)))f_i(\mathbf{\Omega},\iota,\psi)}{r^2/r_0^2}
\end{multline}
As found earlier, for a constant mass ratio, the SNR has a power law dependency on the total mass. So 
\begin{multline}
    E(m_1,m_2,r,\mathbf{\Omega},\iota)\\=\frac{\sum_iE_{0,i}(m_1,m_2)(1+z(r))^{\alpha(m_2/m_1)}f_i(\mathbf{\Omega},\iota,\psi)}{r^2/r_0^2}
\end{multline}

The radial distribution of the sources and the relationship between the luminosity distance ($r$) and redshift ($z$) are given as
\begin{subequations}
\begin{equation}
    P(r)=\frac{\mathcal{R}(r)r^2}{(1+z(r))^{-4}N},\ r<r_{max}
    \label{eq:r}
\end{equation}
\begin{equation}
    r=\frac{c}{H_0}(1+z)\int_0^z((1+x)^3\Omega_m+\Omega_{\Lambda})^{-1/2}dx
    \label{eq:zr}
\end{equation}
\end{subequations}
where maximum luminosity distance $r_{max}$ is assumed for the sources which is well beyond the observation horizon of the detector. $N$ is a normalization constant which may not have an analytical expression. Factor of $(1+z(r))^{-4}$ in Eq. \eqref{eq:r} accounts for the source density dilution and event rate suppression due to cosmological redshift. $\mathcal{R}(r)$ represents the evolution of the local merger rate over the distance. Eq. \eqref{eq:zr} gives the relation between the luminosity distance and the cosmological redshift. $c$ is the speed of light, $H_0=67.6\ {\rm km s^{-1} Mpc^{-1}}$ is the Hubble constant, $\Omega_m=0.31$ and $\Omega_{\Lambda}=0.69$ are the local energy density parameters of matter and cosmological constant (2018 estimates of \emph{Planck} \citep{planck2018}). The effect of radiation density in the computation is neglected as at the related redshifts ($z\lesssim \mathcal{O}(1)$) it contributes negligibly. The universe is assumed to be flat.

Due to the complicated form of $E$ with additional coupling of mass ratio and redshift, the calculations cannot be simplified more analytically, unlike for the static universe. All of the properties including the detection threshold remain coupled and affects the result.

\begin{figure*}[ht]
\gridline{
    \fig{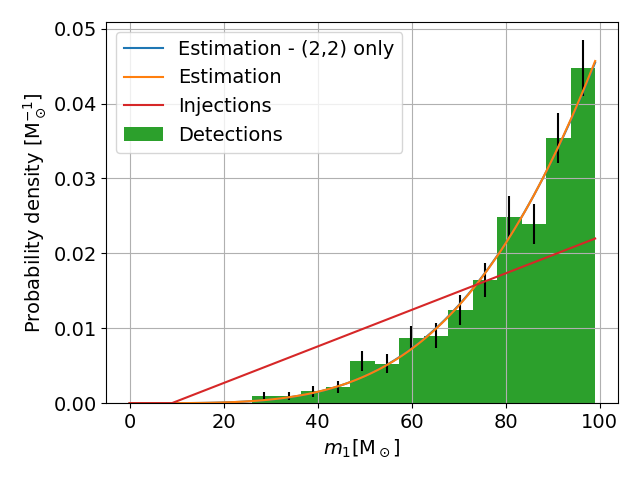}{\columnwidth}{(a)}
    \fig{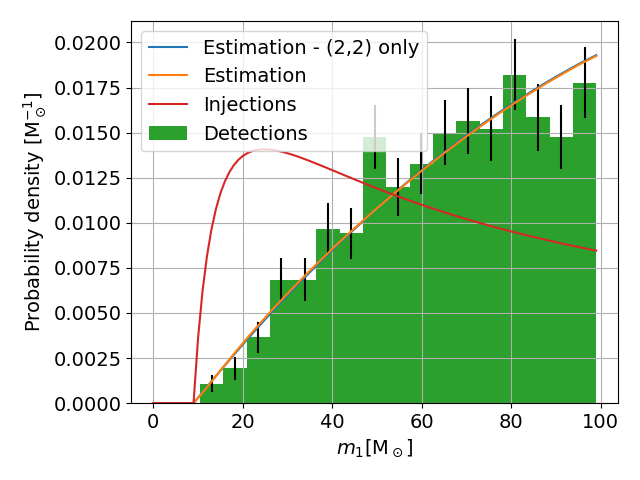}{\columnwidth}{(b)}}
    \caption{Comparison of the semi-analytical estimation and the result of the injection study for the $m_1$ distribution of the observed mergers (a) in the absence of redshift, by using the simplified relation in Eq. \eqref{eq:biasm1}, (b) in the presence of redshift. The black error bars represent one standard deviation of statistical uncertainty. Comparisons with the simulations were done with different mass distributions in order to show the general applicability of the method. The higher order modes are observed to contribute negligibly. Only the simulated distribution with IMRPhenomPv2 waveforms is shown in panel (b). Distribution for the SEOBNRv4 waveforms agrees with the shown distribution within the uncertainties.}
    \label{fig:hist2}
\end{figure*}

The accuracy of our calculation is demonstrated with a simulation study with \textsc{ligo.skymap}. Using the same cosmological estimates, BBH masses in the $(m_1,m_2)$=[10,100]M$_\odot\times$[10,100]M$_\odot$ space were simulated. The mass distribution was chosen to be proportional to the reciprocal of the masses $(P(m_1,m_2)\propto (m_1m_2)^{-1})$. The distribution of $\mathbf{\Omega}$, $\iota$ and $\psi$ are taken such that sources are distributed uniformly on the sky and the orbital orientation direction of the binaries are uniformly distributed. The local merger rate is assumed to be constant. The mergers were detected with one LIGO detector at the same sensitivity used before. The detection threshold was chosen to be $\rho^2_{th}=64$. Maximum luminosity distance was chosen as 30 Gpc which is beyond the maximum detectable distance for the considered masses and the detector configuration. Simulations were done with IMRPhenomPv2 and SEOBNRv4 waveforms separately which agreed each other within the statistical uncertainties. Similarly to the no redshift case, an estimation only assuming the dominant mode was made which showed similar results and yielded the same conclusions. The comparison between the estimation and the results of the simulation study can be seen in Fig. \ref{fig:hist2}. The estimation agrees with the result of the simulation study well which verifies the applicability of the empirical power law dependency of SNR on the total mass alongside with other approximations.

\section{Unmodelled inference of binary black hole population properties}
\label{sec:cosmo}

In Sec. \ref{sec:finding}, the observation bias was analyzed from the point of view of a known astrophysical mass distribution of BBHs and an estimation of the distribution of the masses of detected BBH mergers. In this section, the observation bias is used from the reverse point of view; for making astrophysical inferences from the observations of 46 BBH mergers in GWTC-1 and GWTC-2. These inferences were made by assuming a histogram type astrophysical distribution, similar to \cite{Mandel_2016}; i.e. an unmodelled distribution instead of a functional parametrization such as a power-law. This unmodelled approach allows the inferences to show new structures that are not included in the current models.

The population inference was made via the Bayesian hierarchical inference. This method assigns probabilities to different distributions according to the parameter estimations from measurements \citep{PhysRevD.81.084029}. The inference was made in the ($m_1\geq m_2$) space on $10^{[0.7,2.2]}$M$_\odot\times10^{[0.7,2.2]}$M$_\odot$. In order to have higher resolution for small masses without increasing the total bin count, logarithmic bin sizes were used with 19 bins at each dimension. The prior distribution was chosen to be uniform on the space of distributions for linear masses with logarithmic bin sizes. Further details of computation and discussion on the method are provided in Appendix \ref{ap:pop}.

Fig. \ref{fig:estimate} shows the outcomes of the estimates. Panel (a) shows the ratio of the mean posterior to the mean prior. Two peaks around (34,28) M$_\odot$ and (12,10) M$_\odot$ are seen where the mean posterior is 3.3 and 2.0 times the mean prior distribution. Panel (b) of Fig. \ref{fig:estimate} shows the distribution along $m_1=1.2m_2$ line on which these two peaks lie. It is seen that the peak around (34,28) M$_\odot$ is outside the central 90\% credible region of the effective prior, and the peak around (12,10) M$_\odot$ is outside the central 50\% credible region. There is another peak observed around (18,15) M$_\odot$ although it is not that much significant and lies in the central 50\% credible region of the prior. Corresponding similar structures in the chirp mass distribution were also pointed out in \cite{tiwari2020emergence}. A feature (a peak or a power-law breaking point) around $m_1=33.5$ M$_\odot$ was significantly inferred by \cite{theligoscientificcollaboration2021population} as well. The final observation is the sharper decrease of the posterior mean at heavy masses. This can be observed from panel (a) where for high masses the ratio of the mean posterior to the mean prior reaches down to 1/3. This may be interpreted as a lack of BHs heavier than $\sim40-60$ M$_\odot$ which can be an indication of the predicted pair-instability mass gap. \citep{Woosley_2017,woosley2021pairinstability}.

\begin{figure*}[]
\gridline{
    \fig{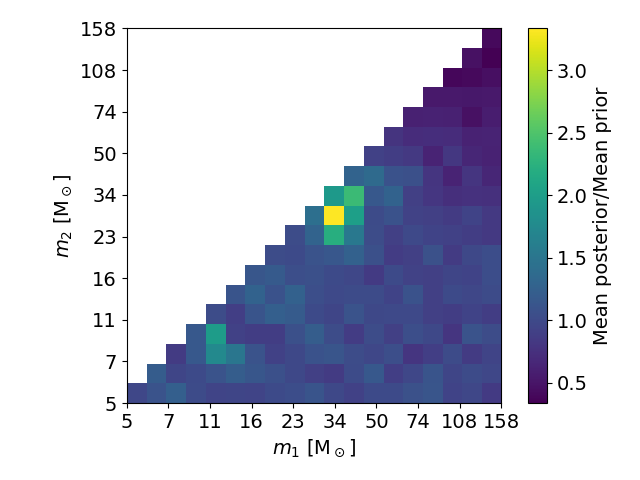}{0.49\textwidth}{(a)}
    \fig{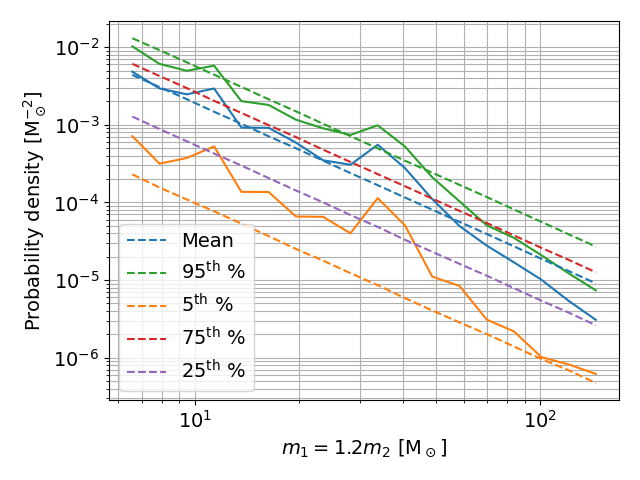}{0.49\textwidth}{(b)}}
\caption{(a) Ratio of the mean posterior distribution to the mean prior distribution (b) Mean probability densities and distributions' percentiles as the bounds of the central credible regions along the $m_1=1.2m_2$ line. Solid and dashed lines represent posterior and prior distributions respectively.}
\label{fig:estimate}
\end{figure*}

\section{Conclusion}
\label{sec:conc}
In this letter, the observation bias in gravitational-wave detections was investigated for non-spinning black holes. By explaining the fundamental origin of the bias, analytical expressions of SNR and source properties were derived. 
By using a numerically computed list of SNRs as a function of $m_1$ and $m_2$, these expressions were evaluated and the agreement with the results from traditional simulations was verified. The advantage of using this semi-analytical method is mainly the reduction of the computational cost; resulting in faster, efficient and more precise estimations. With this algorithm, computations equivalent to $\mathcal{O}(10^{10}$) realizations can be done in $\mathcal{O}(1)$ hours with an average commercial central processing unit core with the processing speed $\mathcal{O}(1)$ GHz. For comparison, the injection campaign described in \cite{theligoscientificcollaboration2021population} has $\mathcal{O}(10^{8}$) realizations which is assumed to have been performed in dedicated computing clusters over longer timescales. Conservatively, it is estimated that the computation of the observation bias can be done $10^6$ times faster with this method than doing it with traditional simulations. Example ready-to-use scripts and some produced datasets for this method are shared in the online repository  \cite{doga_veske_2021_5128048}.

Applying the developed method, unmodelled estimations for the populations of BBHs in GWTC-1 and GWTC-2 were carried out. Excesses of BHs around the mass regions $\sim10$ M$_\odot$ and $\sim30$ M$_\odot$ were observed. Local mean posterior densities around these points lie outside of the 50\% and 90\% credible region of the effective prior while being 3.3 and 2.0 times the mean prior density respectively. Hints of lesser structures and lack of BHs heavier than $\sim40-60$ M$_\odot$ were also observed. With the increasing number of detections, more accurate estimations can be made with more significant structures.

This study concentrated on the bias originating from and effecting the mass distributions while assuming non-spinning black holes; similar to the bias accounting in \cite{theligoscientificcollaboration2021population}. Therefore the differences between the estimates done here and there cannot be originating from the neglection of spin. Any work on spinning black holes is left for future study.

\section*{Acknowledgements}
The authors thank the anonymous referee for guidance, Vaibhav Tiwari for useful feedback and Gayathri V. for comments. This document was reviewed by the LIGO Scientific Collaboration under the document number P2100183.

The authors are grateful for the support of the Columbia University in the City of New York and University of Florida. The Columbia Experimental Gravity group is grateful for the generous support of the National Science Foundation under grant PHY-2012035. I.B. acknowledges the support of the National Science Foundation under grant PHY-1911796 and the Alfred P. Sloan Foundation.
This research has made use of data, software and/or web tools obtained from the Gravitational Wave Open Science Center (\url{https://www.gw-openscience.org/}), a service of LIGO Laboratory, the LIGO Scientific Collaboration and the Virgo Collaboration. LIGO Laboratory and Advanced LIGO are funded by the United States National Science Foundation (NSF) as well as the Science and Technology Facilities Council (STFC) of the United Kingdom, the Max-Planck-Society (MPS), and the State of Niedersachsen/Germany for support of the construction of Advanced LIGO and construction and operation of the GEO600 detector. Additional support for Advanced LIGO was provided by the Australian Research Council. Virgo is funded, through the European Gravitational Observatory (EGO), by the French Centre National de Recherche Scientifique (CNRS), the Italian Istituto Nazionale della Fisica Nucleare (INFN) and the Dutch Nikhef, with contributions by institutions from Belgium, Germany, Greece, Hungary, Ireland, Japan, Monaco, Poland, Portugal, Spain \citep{RICHABBOTT2021100658}.

\bibliography{references}{}

\begin{thebibliography}{}
\expandafter\ifx\csname natexlab\endcsname\relax\def\natexlab#1{#1}\fi
\providecommand{\url}[1]{\href{#1}{#1}}
\providecommand{\dodoi}[1]{doi:~\href{http://doi.org/#1}{\nolinkurl{#1}}}
\providecommand{\doeprint}[1]{\href{http://ascl.net/#1}{\nolinkurl{http://ascl.net/#1}}}
\providecommand{\doarXiv}[1]{\href{https://arxiv.org/abs/#1}{\nolinkurl{https://arxiv.org/abs/#1}}}

\bibitem[{Aasi {et~al.}(2015)Aasi, Abbott, Abbott, Abbott, Abernathy, Ackley,
  Adams, Adams, Addesso, Adhikari, {et~al.}}]{2015}
Aasi, J., Abbott, B.~P., Abbott, R., {et~al.} 2015, Classical and Quantum
  Gravity, 32, 074001, \dodoi{10.1088/0264-9381/32/7/074001}

\bibitem[{Abbott {et~al.}(2019{\natexlab{a}})Abbott, Abbott, Abbott, Abraham,
  Acernese, Ackley, Adams, Adhikari, Adya, Affeldt, \& et~al.}]{Abbott_2019_1}
Abbott, B., Abbott, R., Abbott, T., {et~al.} 2019{\natexlab{a}}, Physical
  Review X, 9, \dodoi{10.1103/physrevx.9.031040}

\bibitem[{Abbott {et~al.}(2016)Abbott, Abbott, Abbott, Abernathy, Acernese,
  Ackley, Adams, Adams, Addesso, Adhikari, {et~al.}}]{PhysRevLett.116.061102}
Abbott, B.~P., Abbott, R., Abbott, T.~D., {et~al.} 2016, Phys. Rev. Lett., 116,
  061102, \dodoi{10.1103/PhysRevLett.116.061102}

\bibitem[{Abbott {et~al.}(2019{\natexlab{b}})Abbott, Abbott, Abbott, Abraham,
  Acernese, Ackley, Adams, Adhikari, Adya, Affeldt, {et~al.}}]{Abbott_2019}
---. 2019{\natexlab{b}}, The Astrophysical Journal, 882, L24,
  \dodoi{10.3847/2041-8213/ab3800}

\bibitem[{Abbott {et~al.}(2020)Abbott, Abbott, Abbott, Abraham, Acernese,
  Ackley, Adams, Adya, Affeldt, Agathos, {et~al.}}]{Abbott_2020}
---. 2020, Classical and Quantum Gravity, 37, 055002,
  \dodoi{10.1088/1361-6382/ab685e}

\bibitem[{Abbott {et~al.}(2021{\natexlab{a}})Abbott, Abbott, Abraham, Acernese,
  Ackley, Adams, Adams, Adhikari, Adya, Affeldt, {et~al.}}]{abbott2021gwtc2}
Abbott, R., Abbott, T.~D., Abraham, S., {et~al.} 2021{\natexlab{a}}, Phys. Rev.
  X, 11, 021053, \dodoi{10.1103/PhysRevX.11.021053}

\bibitem[{Abbott {et~al.}(2021{\natexlab{b}})Abbott, Abbott, Abraham, Acernese,
  Ackley, Adams, Adams, Adhikari, Adya, Affeldt,
  {et~al.}}]{theligoscientificcollaboration2021population}
---. 2021{\natexlab{b}}, The Astrophysical Journal Letters, 913, L7,
  \dodoi{10.3847/2041-8213/abe949}

\bibitem[{Abbott {et~al.}(2021{\natexlab{c}})Abbott, Abbott, Abraham, Acernese,
  Ackley, Adams, Adhikari, Adya, Affeldt, Agathos,
  {et~al.}}]{RICHABBOTT2021100658}
---. 2021{\natexlab{c}}, SoftwareX, 13, 100658,
  \dodoi{https://doi.org/10.1016/j.softx.2021.100658}

\bibitem[{Acernese {et~al.}(2014)Acernese, Agathos, Agatsuma, Aisa, Allemandou,
  Allocca, Amarni, Astone, Balestri, Ballardin, {et~al.}}]{Acernese_2014}
Acernese, F., Agathos, M., Agatsuma, K., {et~al.} 2014, Classical and Quantum
  Gravity, 32, 024001, \dodoi{10.1088/0264-9381/32/2/024001}

\bibitem[{Aghanim {et~al.}(2020)Aghanim, Akrami, Ashdown, Aumont, Baccigalupi,
  Ballardini, Banday, Barreiro, Bartolo, {et~al.}}]{planck2018}
Aghanim, N., Akrami, Y., Ashdown, M., {et~al.} 2020, Astronomy \& Astrophysics,
  641, A6, \dodoi{10.1051/0004-6361/201833910}

\bibitem[{Allen(2005)}]{PhysRevD.71.062001}
Allen, B. 2005, Phys. Rev. D, 71, 062001, \dodoi{10.1103/PhysRevD.71.062001}

\bibitem[{Allen {et~al.}(2012)Allen, Anderson, Brady, Brown, \&
  Creighton}]{Allen_2012}
Allen, B., Anderson, W.~G., Brady, P.~R., Brown, D.~A., \& Creighton, J. D.~E.
  2012, Physical Review D, 85, \dodoi{10.1103/physrevd.85.122006}

\bibitem[{Barausse {et~al.}(2012)Barausse, Morozova, \&
  Rezzolla}]{Barausse_2012}
Barausse, E., Morozova, V., \& Rezzolla, L. 2012, The Astrophysical Journal,
  758, 63, \dodoi{10.1088/0004-637x/758/1/63}

\bibitem[{Bird {et~al.}(2016)Bird, Cholis, Mu\~noz, Ali-Ha\"{\i}moud,
  Kamionkowski, Kovetz, Raccanelli, \& Riess}]{PhysRevLett.116.201301}
Bird, S., Cholis, I., Mu\~noz, J.~B., {et~al.} 2016, Phys. Rev. Lett., 116,
  201301, \dodoi{10.1103/PhysRevLett.116.201301}

\bibitem[{Boh\'e {et~al.}(2017)Boh\'e, Shao, Taracchini, Buonanno, Babak,
  Harry, Hinder, Ossokine, P\"urrer, Raymond, Chu, Fong, Kumar, Pfeiffer,
  Boyle, Hemberger, Kidder, Lovelace, Scheel, \&
  Szil\'agyi}]{PhysRevD.95.044028}
Boh\'e, A., Shao, L., Taracchini, A., {et~al.} 2017, Phys. Rev. D, 95, 044028,
  \dodoi{10.1103/PhysRevD.95.044028}

\bibitem[{Bustillo {et~al.}(2021)Bustillo, Sanchis-Gual, Torres-Forn\'e, Font,
  Vajpeyi, Smith, Herdeiro, Radu, \& Leong}]{PhysRevLett.126.081101}
Bustillo, J.~C., Sanchis-Gual, N., Torres-Forn\'e, A., {et~al.} 2021, Phys.
  Rev. Lett., 126, 081101, \dodoi{10.1103/PhysRevLett.126.081101}

\bibitem[{Calder\'on~Bustillo {et~al.}(2016)Calder\'on~Bustillo, Husa, Sintes,
  \& P\"urrer}]{PhysRevD.93.084019}
Calder\'on~Bustillo, J., Husa, S., Sintes, A.~M., \& P\"urrer, M. 2016, Phys.
  Rev. D, 93, 084019, \dodoi{10.1103/PhysRevD.93.084019}

\bibitem[{Canton \& Harry(2017)}]{canton2017designing}
Canton, T.~D., \& Harry, I.~W. 2017, Designing a template bank to observe
  compact binary coalescences in Advanced LIGO's second observing run.
\newblock \doarXiv{1705.01845}

\bibitem[{Clesse \& García-Bellido(2017)}]{CLESSE2017142}
Clesse, S., \& García-Bellido, J. 2017, Physics of the Dark Universe, 15, 142,
  \dodoi{https://doi.org/10.1016/j.dark.2016.10.002}

\bibitem[{Couch(2012)}]{leon}
Couch, L.~W. 2012, Digital and Analog Communication Systems, 8th edn. (Pearson)

\bibitem[{Field {et~al.}(2014)Field, Galley, Hesthaven, Kaye, \&
  Tiglio}]{PhysRevX.4.031006}
Field, S.~E., Galley, C.~R., Hesthaven, J.~S., Kaye, J., \& Tiglio, M. 2014,
  Phys. Rev. X, 4, 031006, \dodoi{10.1103/PhysRevX.4.031006}

\bibitem[{Finn \& Chernoff(1993)}]{PhysRevD.47.2198}
Finn, L.~S., \& Chernoff, D.~F. 1993, Phys. Rev. D, 47, 2198,
  \dodoi{10.1103/PhysRevD.47.2198}

\bibitem[{Gayathri {et~al.}(2020)Gayathri, Bartos, Haiman, Klimenko, Kocsis,
  M{\'{a}}rka, \& Yang}]{Gayathri_2020}
Gayathri, V., Bartos, I., Haiman, Z., {et~al.} 2020, The Astrophysical Journal,
  890, L20, \dodoi{10.3847/2041-8213/ab745d}

\bibitem[{Gerardi {et~al.}(2021)Gerardi, Feeney, \&
  Alsing}]{gerardi2021unbiased}
Gerardi, F., Feeney, S.~M., \& Alsing, J. 2021.
\newblock \doarXiv{2104.02728}

\bibitem[{Gerosa {et~al.}(2020)Gerosa, Pratten, \& Vecchio}]{Gerosa:2020pgy}
Gerosa, D., Pratten, G., \& Vecchio, A. 2020, Phys. Rev. D, 102, 103020,
  \dodoi{10.1103/PhysRevD.102.103020}

\bibitem[{Katz \& Larson(2018)}]{10.1093/mnras/sty3321}
Katz, M.~L., \& Larson, S.~L. 2018, Monthly Notices of the Royal Astronomical
  Society, 483, 3108, \dodoi{10.1093/mnras/sty3321}

\bibitem[{Kimball {et~al.}(2021)Kimball, Talbot, Berry, Zevin, Thrane,
  Kalogera, Buscicchio, Carney, Dent, Middleton, Payne, Veitch, \&
  Williams}]{kimball2020evidence}
Kimball, C., Talbot, C., Berry, C. P.~L., {et~al.} 2021, The Astrophysical
  Journal Letters, 915, L35, \dodoi{10.3847/2041-8213/ac0aef}

\bibitem[{{Malmquist}(1922)}]{1922MeLuF.100....1M}
{Malmquist}, K.~G. 1922, Meddelanden fran Lunds Astronomiska Observatorium
  Serie I, 100, 1

\bibitem[{Mandel(2010)}]{PhysRevD.81.084029}
Mandel, I. 2010, Phys. Rev. D, 81, 084029, \dodoi{10.1103/PhysRevD.81.084029}

\bibitem[{Mandel {et~al.}(2016)Mandel, Farr, Colonna, Stevenson, Tiňo, \&
  Veitch}]{Mandel_2016}
Mandel, I., Farr, W.~M., Colonna, A., {et~al.} 2016, Monthly Notices of the
  Royal Astronomical Society, 465, 3254–3260, \dodoi{10.1093/mnras/stw2883}

\bibitem[{Mandel {et~al.}(2019)Mandel, Farr, \& Gair}]{Mandel_2019}
Mandel, I., Farr, W.~M., \& Gair, J.~R. 2019, Monthly Notices of the Royal
  Astronomical Society, 486, 1086–1093, \dodoi{10.1093/mnras/stz896}

\bibitem[{Metropolis {et~al.}(1953)Metropolis, Rosenbluth, Rosenbluth, Teller,
  \& Teller}]{doi:10.1063/1.1699114}
Metropolis, N., Rosenbluth, A.~W., Rosenbluth, M.~N., Teller, A.~H., \& Teller,
  E. 1953, The Journal of Chemical Physics, 21, 1087, \dodoi{10.1063/1.1699114}

\bibitem[{Mortlock {et~al.}(2019)Mortlock, Feeney, Peiris, Williamson, \&
  Nissanke}]{PhysRevD.100.103523}
Mortlock, D.~J., Feeney, S.~M., Peiris, H.~V., Williamson, A.~R., \& Nissanke,
  S.~M. 2019, Phys. Rev. D, 100, 103523, \dodoi{10.1103/PhysRevD.100.103523}

\bibitem[{Rodriguez {et~al.}(2016)Rodriguez, Chatterjee, \&
  Rasio}]{PhysRevD.93.084029}
Rodriguez, C.~L., Chatterjee, S., \& Rasio, F.~A. 2016, Phys. Rev. D, 93,
  084029, \dodoi{10.1103/PhysRevD.93.084029}

\bibitem[{Roy {et~al.}(2019)Roy, Sengupta, \& Ajith}]{PhysRevD.99.024048}
Roy, S., Sengupta, A.~S., \& Ajith, P. 2019, Phys. Rev. D, 99, 024048,
  \dodoi{10.1103/PhysRevD.99.024048}

\bibitem[{Roy {et~al.}(2017)Roy, Sengupta, \& Thakor}]{PhysRevD.95.104045}
Roy, S., Sengupta, A.~S., \& Thakor, N. 2017, Phys. Rev. D, 95, 104045,
  \dodoi{10.1103/PhysRevD.95.104045}

\bibitem[{Schutz(2011)}]{Schutz_2011}
Schutz, B.~F. 2011, Classical and Quantum Gravity, 28, 125023,
  \dodoi{10.1088/0264-9381/28/12/125023}

\bibitem[{{Talbot} \& {Thrane}(2020)}]{2020arXiv201201317T}
{Talbot}, C., \& {Thrane}, E. 2020, arXiv e-prints, arXiv:2012.01317.
\newblock \doarXiv{2012.01317}

\bibitem[{Thrane \& Talbot(2020)}]{thrane_talbot_2020}
Thrane, E., \& Talbot, C. 2020, Publications of the Astronomical Society of
  Australia, 37, e036, \dodoi{10.1017/pasa.2020.23}

\bibitem[{Tiwari \& Fairhurst(2021)}]{tiwari2020emergence}
Tiwari, V., \& Fairhurst, S. 2021, The Astrophysical Journal Letters, 913, L19,
  \dodoi{10.3847/2041-8213/abfbe7}

\bibitem[{Usman {et~al.}(2016)Usman, Nitz, Harry, Biwer, Brown, Cabero, Capano,
  Canton, Dent, Fairhurst, Kehl, Keppel, Krishnan, Lenon, Lundgren, Nielsen,
  Pekowsky, Pfeiffer, Saulson, West, \& Willis}]{Usman_2016}
Usman, S.~A., Nitz, A.~H., Harry, I.~W., {et~al.} 2016, Classical and Quantum
  Gravity, 33, 215004, \dodoi{10.1088/0264-9381/33/21/215004}

\bibitem[{Varma {et~al.}(2019)Varma, Field, Scheel, Blackman, Kidder, \&
  Pfeiffer}]{PhysRevD.99.064045}
Varma, V., Field, S.~E., Scheel, M.~A., {et~al.} 2019, Phys. Rev. D, 99,
  064045, \dodoi{10.1103/PhysRevD.99.064045}

\bibitem[{Veske(2021)}]{doga_veske_2021_5128048}
Veske, D. 2021, dveske/observation-bias-gw: GW observation bias, v1.1,  Zenodo,
  \dodoi{10.5281/zenodo.5128048}

\bibitem[{Veske {et~al.}(2020)Veske, Márka, Sullivan, Bartos, Rainer Corley,
  Samsing, \& Márka}]{10.1093/mnrasl/slaa123}
Veske, D., Márka, Z., Sullivan, A.~G., {et~al.} 2020, Monthly Notices of the
  Royal Astronomical Society: Letters, 498, L46, \dodoi{10.1093/mnrasl/slaa123}

\bibitem[{Veske {et~al.}(2021)Veske, Sullivan, M{\'{a}}rka, Bartos, Corley,
  Samsing, Buscicchio, \& M{\'{a}}rka}]{Veske_2021}
Veske, D., Sullivan, A.~G., M{\'{a}}rka, Z., {et~al.} 2021, The Astrophysical
  Journal Letters, 907, L48, \dodoi{10.3847/2041-8213/abd721}

\bibitem[{Woosley(2017)}]{Woosley_2017}
Woosley, S.~E. 2017, The Astrophysical Journal, 836, 244,
  \dodoi{10.3847/1538-4357/836/2/244}

\bibitem[{Woosley \& Heger(2021)}]{woosley2021pairinstability}
Woosley, S.~E., \& Heger, A. 2021, The Astrophysical Journal Letters, 912, L31,
  \dodoi{10.3847/2041-8213/abf2c4}

\bibitem[{Yang {et~al.}(2019)Yang, Bartos, Gayathri, Ford, Haiman, Klimenko,
  Kocsis, M\'arka, M\'arka, McKernan, \&
  O'Shaughnessy}]{PhysRevLett.123.181101}
Yang, Y., Bartos, I., Gayathri, V., {et~al.} 2019, Phys. Rev. Lett., 123,
  181101, \dodoi{10.1103/PhysRevLett.123.181101}

\bibitem[{Zevin {et~al.}(2021)Zevin, Bavera, Berry, Kalogera, Fragos, Marchant,
  Rodriguez, Antonini, Holz, \& Pankow}]{Zevin_2021}
Zevin, M., Bavera, S.~S., Berry, C. P.~L., {et~al.} 2021, The Astrophysical
  Journal, 910, 152, \dodoi{10.3847/1538-4357/abe40e}

\end{thebibliography}
\bibliographystyle{aasjournal}

\appendix
\section{Basics on gravitational-wave detection}

\subsection{Detected inspiral waveform}
\label{sec:gw}
The observed strain in a detector network can be written by decomposing the contributions from two polarizations $h_+$ and $h_\times$ as
\begin{equation}
    h(t)=h_+(t)F_+(\mathbf{\Omega},\psi)+h_\times(t)F_\times(\mathbf{\Omega},\psi)
\end{equation}
where $F_+$ and $F_\times$ are the antenna patterns of the detector network for the two polarizations. For a single two-armed interferometric detector with $90^\circ$ angle between its arms, the antenna patterns are given as \citep{Schutz_2011}
\begin{subequations}
    \begin{equation}
        F_+ = \frac{1}{2}(1+cos^2\delta)cos2\theta cos2\psi-cos\delta sin2\theta sin2\psi
    \end{equation}
    \begin{equation}
        F_\times = \frac{1}{2}(1+cos^2\delta)cos2\theta sin2\psi+cos\delta sin2\theta cos2\psi
    \end{equation}
\end{subequations}
where $\mathbf{\Omega}=(\delta,\theta)$ are the zenith (measured from z axis to xy plane) and azimuth (measured from x axis to y axis) angles in a detector centered coordinate system where detector's arms lie along x and y axes. $\psi$ is the rotational angle between the x axis of the detector centered coordinate system and the projection of the x axis of the coordinate system where $h_+$ and $h_\times$ are defined (radiation frame) to the detector's plane.

For two polarizations, the inspiral waveforms for the dominant mode (2,2) in radiation frame are \citep{PhysRevD.47.2198}
\begin{subequations}
    \begin{equation}h_+(t)=2\frac{G^{5/3}\mathcal{M}^{5/3}}{rc^{2/3}}(1+cos^2\iota)(\pi f(t))^{2/3}cos(\phi_0+\Phi(t))\end{equation}
    \begin{equation}h_\times(t)=4\frac{G^{5/3}\mathcal{M}^{5/3}}{rc^{2/3}}cos\iota(\pi f(t))^{2/3}sin(\phi_0+\Phi(t))\end{equation}
\end{subequations}
where chirp mass $\mathcal{M}$ is defined as
\begin{equation}
    \mathcal{M}=\frac{(m_1m_2)^{0.6}}{(m_1+m_2)^{0.2}}
\end{equation}
The frequency of the wave ($f)$ and the accumulated phase ($\Phi$) are given as
\begin{equation}
    f(t)=\frac{1}{\pi}(\frac{c}{G\mathcal{M}})^{5/8}(\frac{5}{256(t_m-t)})^{3/8}
\end{equation}
\begin{equation}
    \Phi(t)=\int_0^t 2\pi f(\tau) d\tau=-2(\frac{c(t_m-t)}{5G\mathcal{M}})^{5/8}
\end{equation}
where $t_m$ is the time of the merger. However, these given inspiral waveforms do not hold up to $t_m$. The actual waveforms starts deviating from these forms as the black holes come closer.
Collecting all the constants under a single constant $C$, $h(t)^2$ can be written as
\begin{multline}
    h(t)^2=C\frac{\mathcal{M}^{5/2}}{(t_m-t)^{1/2}r^2}(F_+^2(\frac{1+cos^2\iota}{2})^2+F_\times^2cos^2\iota)\\ \times cos^2[\phi_0+\Phi(t)+arctan(\frac{2F_\times cos\iota}{F_+(1+cos^2\iota)})]
    \label{eq:h2}
\end{multline}
\subsection{Matched filtering}
\label{sec:mf}
Matched filtering is the optimal method for detecting a signal with a known waveform in the presence of an additive Gaussian noise with a known spectrum \citep{leon}. The filtering maximizes the signal-to-noise ratio (SNR) which is a monotonically increasing function of the likelihood ratio of having the sought signal in the data to having only noise. Consequently, setting an SNR threshold as a detection criteria can be used optimally when the mentioned conditions above are satisfied. The power SNR ($\rho^2$) for a search looking for a real waveform $h(t)$ with unknown amplitude and arrival time, in the noisy data $w(t)=\alpha h(t-t_0)+n(t)$ can be calculated as
\begin{equation}
    \rho^2(t) = \int^\infty_{-\infty} \frac{H^*(f)W(f)}{S_n(f)}e^{-j2\pi f t}df
    \label{eq:mf1}
\end{equation}
where $j=\sqrt{-1}$, $H^*(f)$ is the complex conjugate of the Fourier transform of $h(t)$, $W(f)$ is the Fourier transform of $w(t)$ and $S_n(f)$ is the two-sided power spectral density of the noise $n$. If the noise additionally has a white spectrum, then up to constants, SNR can be calculated in time domain as 
\begin{equation}
    \rho^2(t) \propto \int^\infty_{-\infty} h(\tau-t)w(\tau)d\tau
\end{equation}
The time dependency of SNR represents the delay in the arrival time of the signal with respect to the start of $w(t)$. If $\rho^2(t_d)$ exceeds the predetermined SNR threshold $\rho^2_{th}$, which is based on the allowed false alarm probability of the search, one can claim to have detected the signal with the determined false alarm probability, $t_d$ after the start of the data taking. Since many templates are searched in gravitational-wave searches, in order to have a single threshold, SNR of each template is further normalized with the power of each template, which is explained thoroughly in \cite{Allen_2012}. Although in gravitational-wave searches based on matched filtering the detection threshold is the false-alarm rate but not the bare SNR because of the non-Gaussian Poisson-like noise called glitches, the detections happen in well correlation with SNR especially after glitch involving data parts are removed \citep{Abbott_2020}. The important take away from this subsection is the fact that the mean power SNR increases linearly with the integral of $h^2$.
\subsection{Details on Population Inference}
\label{ap:pop}
Assuming a uniform prior on the distribution space and a Jeffrey's prior (reciprocal) on the event rate which is assumed to be constant and independent of other parameters, the probability of a histogram type distribution $f$ for the variables $\boldsymbol{\lambda}$ (i.e. $m_1$ and $m_2$) can be found up to a constant as

\begin{equation}
P(f) \propto \frac{\prod_{i=1}^{n} \sum_{\boldsymbol{\lambda}} f(\boldsymbol{\lambda}) P_i(\boldsymbol{\lambda})P(D_i|\boldsymbol{\lambda})\Delta(\boldsymbol{\lambda})}{(\sum_{\boldsymbol{\lambda}} f(\boldsymbol{\lambda})\sum_{j=1}^{N} \Delta t_j P(D_j|\boldsymbol{\lambda})\Delta(\boldsymbol{\lambda}))^n}
\end{equation}
where $P_i$ are the parameter estimations from $n$ measurements, $D_i$ represent the detectability for $N$ different networks of gravitational-wave detectors and $\Delta t_j$ are the cumulative operation time of each network. Here a detector network is defined by the detectors in it as well as its detectors' sensitivities. Networks composed of the same detectors during different observing runs are considered as different networks. $\Delta(\boldsymbol{\lambda})$ are the bin sizes.

Using the SNR of the weakest signals in each catalog, the detection threshold for O3a networks were chosen as amplitude SNR=8 and for O1-O2 networks as amplitude SNR=10. Due to the non-Gaussian noise in the detectors, generally a down-scaled SNR is used as a ranking statistic after a chi-squared test \citep{PhysRevD.71.062001,Usman_2016}. This test determines whether the time-frequency distribution of the power in the detected signal is consistent with the matching waveform template and penalizes SNR according to the inconsistent power. For astrophysical signals, the down-scaled SNR is expected to be equal or approximately equal to the original SNR. Since confirmed astrophysical detections were considered here, the ranking statistic for them was taken as the SNR directly. For O3 detection probabilities, for networks other than the ones had detections (HLV, HL, LV, L), the probabilities were taken as 0. For SNR calculations, PSDs of GW151012, GW170809 and GW190412 were used for O1, O2 and O3 networks respectively. Parameter estimations of the 46 BBH mergers in GWTC-1 and GWTC-2 were used.

A non-evolving local merger rate that is independent of the mass distribution was assumed. Since hierarchical inference doesn't provide a single distribution but rather assigns probabilities to infinite number of distributions, a metric needs to be chosen to comprehend the general tendencies. Here the mean distribution and bounds of the central 90\% credible regions for each bin were chosen as the metrics. One may desire to find the most probable distribution instead of the mean distribution. However, with finite number of measurements the most probable distribution is guaranteed to be composed of delta functions. These delta functions would be positioned at the highest probability locations of each parameter estimation, except for closely neighbouring estimations which can produce less number of but more strong delta functions between their peaks. Since such distributions physically don't make sense; instead of the most probable distribution, the mean distribution and the credible regions were considered here. Another property this inference method has is; with the decreasing number of measurements and increasing bin count, posterior approaches to prior and becomes not informative. This can be interpreted as; as there are more parameters (here values of bins) to be estimated, the same amount of observation becomes less relevant (or vice versa). In other words, high number of measurements (relative to the total bin count) is required in order to have substantial data driven effects with high resolution. Therefore the mean posterior distribution should be carefully used as an astrophysical distribution as it depends on our choice of bin sizes which is not astrophysical. The reason for using the logarithmic bin sizes here is to keep the total bin count at some level while increasing the resolution for lower masses.

Sampling over the distributions was done via Markov Chain Monte Carlo sampling using Metropolis algorithm \citep{doi:10.1063/1.1699114}. Candidate distributions at each iteration of sampling was chosen independently of the current distribution. Each candidate distribution was first generated from a flat Dirichlet distribution (uniform distribution on the distribution space) for total of 190 bins (171 of 361 bins correspond to $m_1<m_2$). Then value of each bin was rescaled with the inverse of its bin size. Since scaling is an affine transformation, it doesn't modify the density distribution of distributions, i.e. it maps uniform distribution to uniform distribution. Therefore this generation-scaling process is equivalent to generating distributions directly on the space of distributions for linear masses with logarithmic bin sizes with equal probabilities, i.e. uniformly. Total of 4$\times10^6$ iterations were performed. 
\end{document}